\documentclass[doublecol,final]{epl2tq} 

\usepackage{graphics}
\usepackage{bbm}
\usepackage{amsmath}
\usepackage[utf8]{inputenc}

\title{Which-way measurement and momentum kicks}
\author{Tabish Qureshi
\thanks{E-mail: \email{tabish@ctp-jamia.res.in}}}
\shortauthor{Tabish Qureshi}

\institute{Centre for Theoretical Physics, Jamia Millia Islamia, New Delhi,
India.}


\abstract{
The two-slit interference experiment with a which-way detector has been a
topic of intense debate. Scientific community is divided on the question
whether the particle receives a momentum kick because of the process
of which-way measurement. It is shown here that the same experiment can
be viewed in two different ways, depending on which basis of the which-way
detector states one chooses to look at. In one view, the loss of interference
arises due to the entanglement of the two paths of the particle with
two orthogonal states of the which-way detector. In another view, the
loss of interference can be interpreted as arising from random momentum
kicks of magnitude $h/2d$ received by the particle, $d$ being the
slit separation. The same scenario is shown to hold for a three-slit
interference experiment. The random momentum kicks for the three-slit
case are of two kinds, of magnitude $\pm h/3d$. The analysis is also
generalized to the case of n-slit interference.  The two alternate views
are described by the same quantum state, and hence are completely
equivalent. The concept of "local" versus "nonlocal" kicks, much
discussed in the literature, is not needed here. }

\begin{document}

\maketitle

\section{Introduction}

The two-slit experiment with particles has become a cornerstone for the issue
of wave-particle duality or the concept of complementarity introduced by
Niels Bohr \cite{bohr}. A debate was set in motion when Einstein proposed
his famous recoiling-slit experiment, in an unsuccessful bid to refute
Bohr's complementarity principle \cite{recoil}. This thought experiment
has now been beautifully realized in different ways
\cite{recoillui,recoildorner,utter}.  Bohr had countered Einstein by pointing
out that measuring the momentum of the recoiling slit, in order to find
which slit the particle went through, would produce an uncertainty in the
position of the recoiling slit, which in turn would wash out the interference.
Bohr's specific resolution led many authors to surmise that complementarity
was probably another way of stating the uncertainty principle, and that
complementarity has its roots in the uncertainty principle. Scully, Englert
and Walther proposed a which-way experiment using micromaser cavities
which, they claimed, does not involve any position-momentum uncertainty
\cite{SEW}. Their conclusion was that the which-way detection process
does not involve any momentum tranfer to the interfering particle.
Storey et.al. countered this claim by proving that if an interference
pattern is destroyed in a which-way experiment, a momentum of at least the
magnitude $\hbar/d$ should be transfered to the particle, where $d$ is the
separation between the two slits \cite{storey}.
A momentum transfer of an amount smaller than that would not destroy
the interference completely. This led to a shift in the focus of the
debate to the question as to whether there is a momentum transfer to the
particle involved in the process of which-way detection
\cite{kick-englert,kick-storey,kick-wiseman1,kick-wiseman2,kick-wiseman3}.

\begin{figure}
\centerline{\resizebox{8.5cm}{!}{\includegraphics{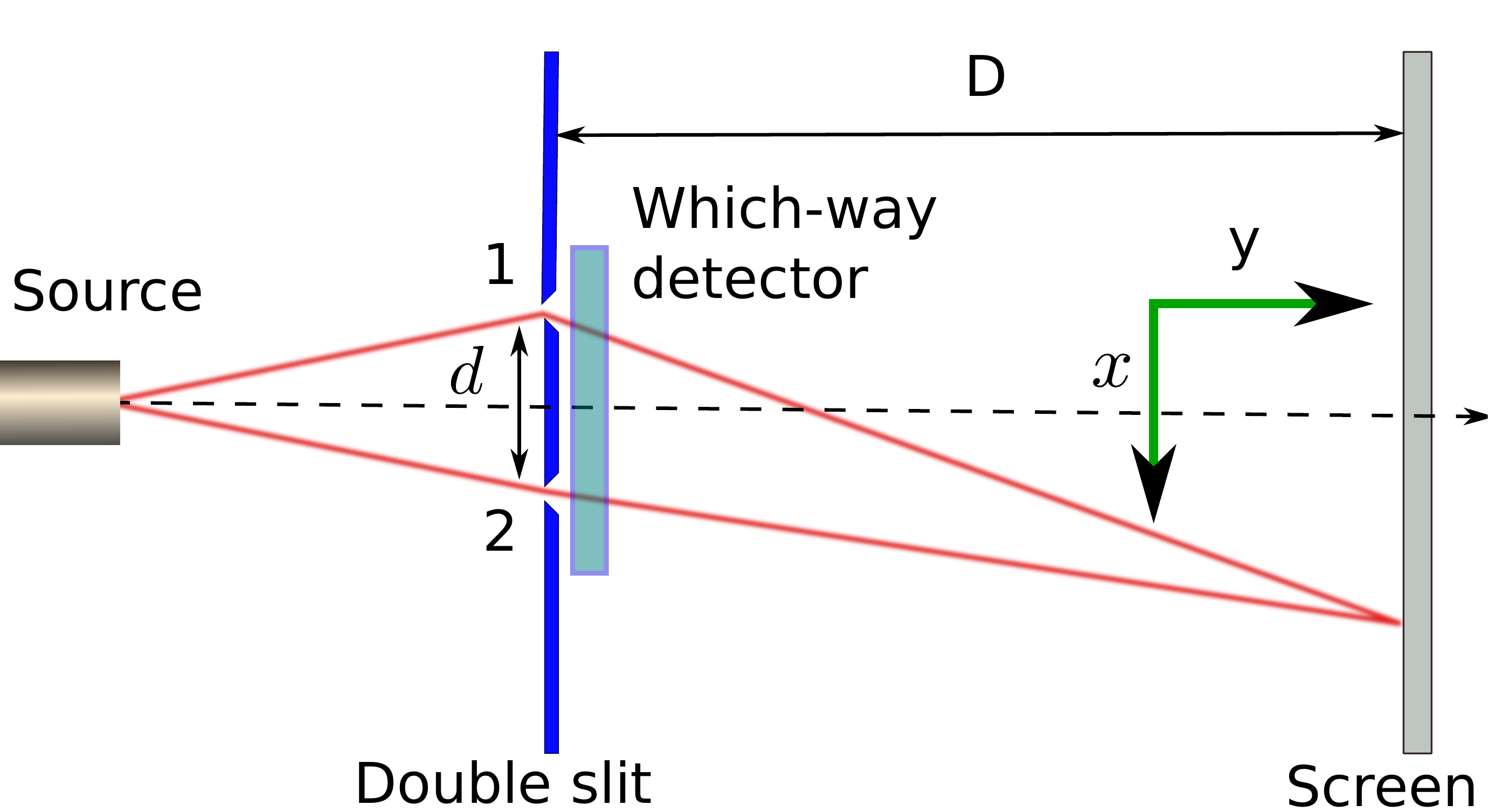}}}
\caption{Schematic diagram of a two-slit interference experiment in the
presence of a which-way detector. Slits 1 and 2 are located at $x=0$
and  $x=d$, respectively.}
\label{twoslit}
\end{figure}

Later it was shown that the complementarity principle can be understood
in terms of the ubiquitous entanglement between the particle and the
which-way detector, and also equivalently in terms of the uncertainty between
certain operators of the which-way detector, and the well-known wave-particle
duality relation \cite{englert} can be derived from both \cite{tqrv,durr}.
However, the question as to 
whether there is a momentum transfer to the particle or not, still appears
to be not settled \cite{mir,xiao}. Wiseman has tried to reconcile between the two views by
introducing the concept of {\em nonlocal} momentum kicks \cite{kick-wiseman2},
and has proposed that the momentum kicks could be probed through weak
measurements \cite{wiseman2}.
In the following
we show that the two views which say that there is, and there isn't a momentum
kick, are completely equivalent.

\section{Two-slit which-way experiment}

Let $\psi(x)$ represent the state of the particle just as it emerges from
the double-slit:
\begin{equation}
\psi(x) = \tfrac{1}{\sqrt{2}} \left[ \psi_1(x) + \psi_2(x) \right],
\label{ent0}
\end{equation}
where $\psi_1,\psi_2$ are narrow states localized at $x=0$ and
$x=d$, respectively. The states $\psi_1,\psi_2$ are orthogonal by virtue
of their spatial separation.
Let us now introduce a which-way detector at the double-slit, as schematically
shown in FIG. \ref{twoslit}. We need not assume any form of the which-way
detector. According to von Neumann's concept of measurement \cite{neumann},
if the which-way detector is to make a measurement on which of the two
paths the particle followed, it's two states should get entangled with the
states of the two paths:
\begin{equation}
\Psi(x) = \tfrac{1}{\sqrt{2}} \left[ \psi_1(x)|d_1\rangle + \psi_2(x)|d_2\rangle \right],
\label{ent1}
\end{equation}
where $|d_1\rangle, |d_2\rangle$ are certain normalized states of the 
which-way detector. The particle travels a distance $D$ to the screen in 
a time $t$, and the state is given by
\begin{equation}
\Psi(x,t) = \tfrac{1}{\sqrt{2}} \left[ \psi_1(x,t)|d_1\rangle + \psi_2(x,t)|d_2\rangle \right],
\label{ent2}
\end{equation}
where $\psi_1(x,t),\psi_2(x,t)$ remain orthogonal.
Now it is easy to see that when one calculates the probability density of
the particle falling on the screen at a position $x$, namely $|\Psi(x,t)|^2$
the cross terms in the above, which represent interference, have a factor
proportional to $|\langle d_1|d_2\rangle|$:
\begin{eqnarray}
|\Psi(x,t)|^2 &=& \tfrac{1}{2} \left[ |\psi_1(t)|^2 + |\psi_2(t)|^2 \right.\nonumber\\
&&\left. + \psi_1^*(t)\psi_2(t)\langle d_1|d_2\rangle + 
\psi_2^*(t)\psi_1(t)\langle d_2|d_1\rangle\right],\nonumber\\
\end{eqnarray}
where we have suppressed the $x$ dependence of the states for brevity.
If $|d_1\rangle,|d_2\rangle$ are orthogonal, the last two terms in the above
equation, which represent interference, drop off.
It needs to be stressed that $\psi_1(x),\psi_2(x)$ in (\ref{ent1}) are the
same as those in (\ref{ent0}), and the which-way detection does not change
the individual states of the particle emerging from the slits,
hence there is no question of any additional
momentum kick due to the which-way detection. This was the point of view
of Scully, Englert and Walther \cite{SEW}.

However, the loss of interference described here may also be viewed in a
slightly different fashion. If $|d_1\rangle,|d_2\rangle$ are orthonormal,
one can introduce another set of orthonormal states:
$|d_{\pm}\rangle =|d_1\rangle \pm |d_2\rangle)/\sqrt{2}$.
The state of the particle emerging from the double-slit and the which-way
detector combined, (\ref{ent1}) can then be written as
\begin{eqnarray}
\Psi(x) &=& \tfrac{1}{2} [ \psi_1(x)+ \psi_2(x)]|d_+\rangle \nonumber\\
&& + \tfrac{1}{2} [ \psi_1(x) - \psi_2(x)]|d_-\rangle.
\label{ent3a}
\end{eqnarray}
The state of the particle at the screen and the which-way detector combined,
(\ref{ent2}) can then be written as
\begin{eqnarray}
\Psi(x,t) &=& \tfrac{1}{2} [ \psi_1(x,t)+ \psi_2(x,t)]|d_+\rangle \nonumber\\
&& + \tfrac{1}{2} [ \psi_1(x,t) - \psi_2(x,t)]|d_-\rangle.
\label{ent3}
\end{eqnarray}
Although (\ref{ent3}) shows no interference, if the particle is detected in
coincidence with the which-way state $|d_+\rangle$, it shows an interference
which is exactly the same as that shown by (\ref{ent0}). Alternately,
if the particle is detected in coincidence with the which-way state
$|d_-\rangle$, it shows an interference which is slightly shifted. The two
interferences may be represented as
\begin{eqnarray}
|\Psi_+(x,t)|^2 &=& \tfrac{1}{4} \left[ |\psi_1(t)|^2 + |\psi_2(t)|^2 \right.\nonumber\\
&&\left. + \psi_1^*(t)\psi_2(t)+ \psi_2^*(t)\psi_1(t)\right],\nonumber\\
|\Psi_-(x,t)|^2 &=& \tfrac{1}{4} \left[ |\psi_1(t)|^2 + |\psi_2(t)|^2 \right.\nonumber\\
&&\left. - \psi_1^*(t)\psi_2(t)- \psi_2^*(t)\psi_1(t)\right],
\end{eqnarray}
where $\Psi_{\pm}(x,t) = \langle d_{\pm}|\Psi(x,t)\rangle$.
A temporary mixing of the Dirac notation may be excused here.
The fact that coincident detection of the particle with $|d_{\pm}\rangle$
states brings back interference, is called quantum erasure \cite{eraser,delayed}.
Looking at (\ref{ent3}), one can understand the loss of interference as
arising due to the spinor $|d_-\rangle$ flipping the relative phase between
the two paths by $\pi$. This has been recognized earlier \cite{luis,unni}.

Now we wish to point out that the phase-flip in (\ref{ent3}) can also
be interpreted as a momentum kick. We write (\ref{ent3a}) as
\begin{eqnarray}
\Psi(x) &=& \tfrac{1}{2} [ \psi_1(x)+ \psi_2(x)]|d_+\rangle \nonumber\\
&& + e^{ip_{0}x/\hbar} \tfrac{1}{2} [ \psi_1(x) + \psi_2(x)]|d_-\rangle,
\label{entkick}
\end{eqnarray}
where $p_0=\hbar\pi/d$ is a momentum-kick the particle receives whenever
the which-way detector state is $|d_-\rangle$. To see if writing (\ref{ent3})
as (\ref{entkick}) is valid or not, we simplify (\ref{entkick}) as
\begin{eqnarray}
\Psi(x) &=& \tfrac{1}{2} [ \psi_1(x)+ \psi_2(x)]|d_+\rangle \nonumber\\
&& + \tfrac{1}{2} [ e^{ip_{0}x/\hbar} \psi_1(x) + e^{ip_{0}x/\hbar} \psi_2(x)]|d_-\rangle \nonumber\\
&=& \tfrac{1}{2} [ \psi_1(x)+ \psi_2(x)]|d_+\rangle \nonumber\\
&& + \tfrac{1}{2} [ \psi_1(x,t) + e^{ip_{0}d/\hbar} \psi_2(x)]|d_-\rangle \nonumber\\
&=& \tfrac{1}{2} [ \psi_1(x)+ \psi_2(x)]|d_+\rangle \nonumber\\
&& + \tfrac{1}{2} [ \psi_1(x) - \psi_2(x)]|d_-\rangle,
\label{entkick1}
\end{eqnarray}
where we have used the fact that $\psi_1(x)$ is a narrow state localized at
$x=0$ and $\psi_2(x)$ is a narrow state localized at $x=d$. This simple
analysis shows that (\ref{entkick}) is the same as (\ref{ent3a}), and
the phase flip can also be seen as momentum kick of magnitude $p_0=h/2d$
which the particle receives randomly, fifty percent of the time, i.e,
whenever the which-way detector state is $|d_-\rangle$.

Now (\ref{ent1}) and (\ref{entkick}) represent the same state. If one 
viewed the which-way detector in the basis described by $|d_1\rangle,|d_2\rangle$,
one would say that the two paths of the particle are correlated with two
orthogonal states, and so there is no interference.
Alternately if one viewed the which-way detector in the basis described by
$|d_+\rangle,|d_-\rangle$, one would say that the particles receives a
momentum kick equal to $p_0=h/2d$, fifty percent of the the time,  and
the two interference patterns corresponding to the particles which receive
or do not receive a kick, are mutually shifted, washing out the result.

\subsection{Choice of basis}

One may wonder what happens if one uses another basis. The answer is that
only a basis whose states are unbiased with respect to $|d_1\rangle, 
|d_2\rangle$, will lead to the concept of momentum kicks. We demonstrate
that in the following analysis. Suppose we choose another basis
$|\alpha\rangle, |\beta\rangle$ for the path-detector, such that
\begin{eqnarray}
	|d_1\rangle &=& \frac{1}{\sqrt{2}} \left( e^{i\theta_1}|\alpha\rangle
	+ e^{i\theta_2}|\beta\rangle \right) \nonumber\\
	|d_2\rangle &=& \frac{1}{\sqrt{2}} \left( e^{i\theta_3}|\alpha\rangle
	+ e^{i\theta_4}|\beta\rangle \right).
\end{eqnarray}
The above represent the most general basis which is unbiased with respect to
$|d_1\rangle, |d_2\rangle$. Orthogonality of $|d_1\rangle,|d_2\rangle$
demands that $\theta_1-\theta_2=\theta_3-\theta_4+\pi$.
The state of the particle at the screen and the which-way detector combined,
(\ref{ent1}) can then be written as
\begin{eqnarray}
\Psi(x) &=& \tfrac{1}{2} (e^{i\theta_1}\psi_1+ e^{i\theta_3}\psi_2)|\alpha\rangle \nonumber\\
&& + \tfrac{1}{2} (e^{i\theta_2}\psi_1 + e^{i\theta_4}\psi_2)|\beta\rangle,
\nonumber\\
  &=& \tfrac{1}{2} (e^{i\theta_1}\psi_1+ e^{i\theta_3}\psi_2)|\alpha\rangle \nonumber\\
	&& + \tfrac{1}{2}e^{i(\theta_2-\theta_1)} e^{ip_{0}x\over\hbar}
	(e^{i\theta_1}\psi_1 + e^{i\theta_3}\psi_2)|\beta\rangle, 
\label{entg}
\end{eqnarray}
where $p_0=h/2d$. Again, we have used the fact that $\psi_1, \psi_2$ are
narrow states localized at $x=0$ and $x=d$, respectively.
In this general case too, the state of the particle, corresponding to the
two path-detector states, is the same except the term $e^{ip_{0}x/\hbar}$,
when the path-detector state is $|\beta\rangle$. Thus, the magnitude
of the momentum kick is the same whatever basis states one chooses.
The phase factor $e^{i(\theta_2-\theta_1)}$ is unimportant because it does not
affect the position of the conditional interference pattern corresponding to
the path-detector state $|\beta\rangle$. If one assumes that the two
slits are located at $x=\pm d/2$, instead of being at $x=0,d$, one gets
an additional phase factor of $e^{i\pi/2}$.
If one chooses to use a path-detector
basis which is not unbiased with respect to $|d_1\rangle, |d_2\rangle$,  
the experiment cannot be interpreted in terms of momentum kicks.

\section{Three-slit which-way experiment}

\begin{figure}
\centerline{\resizebox{8.5cm}{!}{\includegraphics{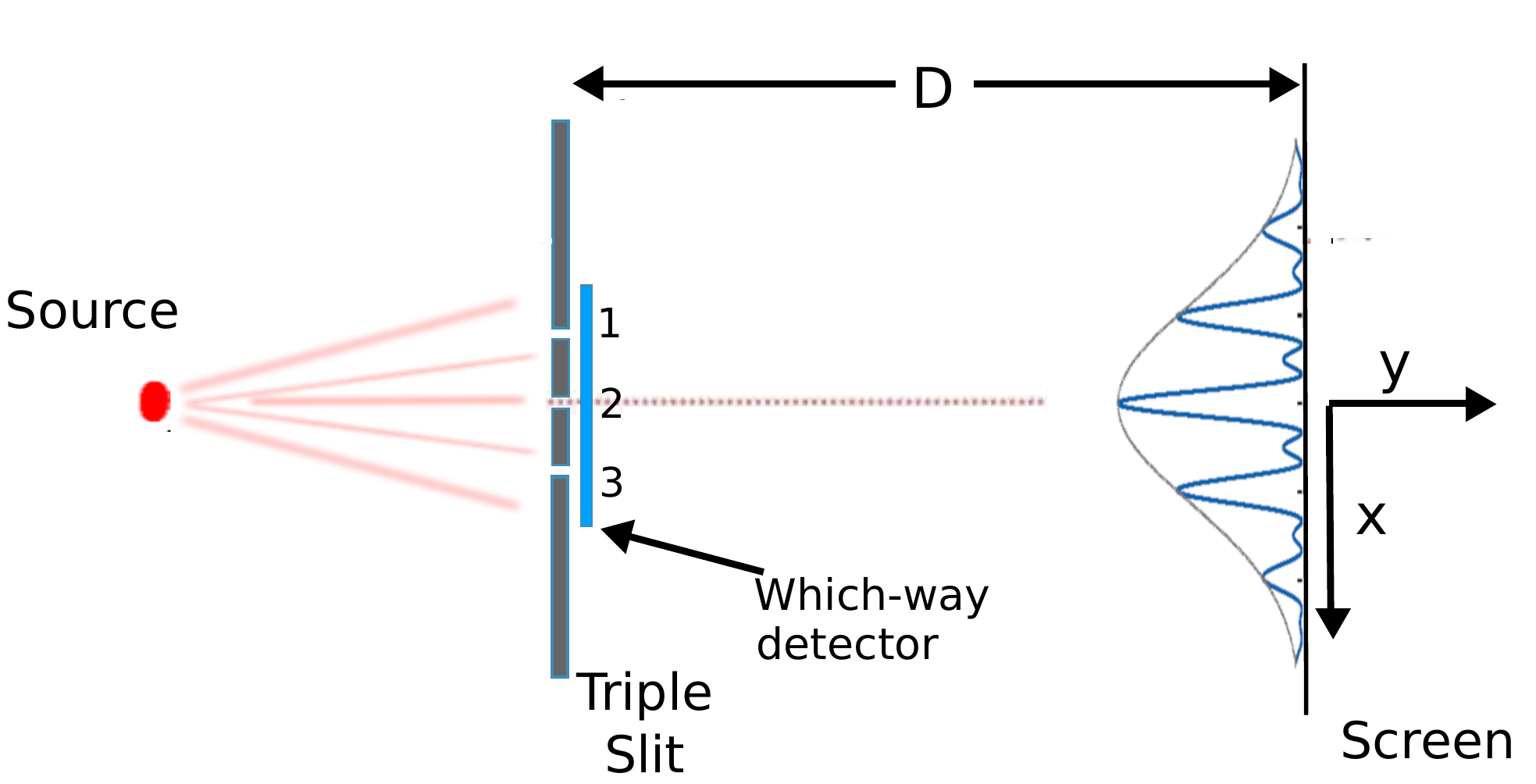}}}
\caption{Schematic diagram of a three-slit interference experiment in the
presence of a which-way detector. Slits 1, 2 and 3 are located at
$x=-d$, $x=0$ and $x=d$, respectively.}
\label{3slit}
\end{figure}

To convince the reader that the above view is not contrived, but very 
natural, we extend it to a three-slit interference experiment in the
presence of a which-way detector (see FIG. \ref{3slit}).
The state of the particle plus the which-way detector, as it emerges from
the triple-slit, can be written as
\begin{equation}
\Psi(x) = \tfrac{1}{\sqrt{3}} \left[ \psi_1(x)|d_1\rangle +
\psi_2(x)|d_2\rangle + \psi_3(x)|d_3\rangle \right],
\label{3ent1}
\end{equation}
where $\psi_1,\psi_2,\psi_3$ are the states corresponding to the three paths
of the particle and $|d_1\rangle,|d_2\rangle,|d_3\rangle$ are the
orthonormal states of the which-way detector corresponding to those paths.
It is obvious that (\ref{3ent1}) will not show any interference because of
the orthogonality of $|d_1\rangle,|d_2\rangle,|d_3\rangle$.

Proceeding along the lines of the preceding discussion, one may
consider a scheme of quantum erasure of three-slit interference \cite{3eraser}
and define three mutually orthogonal states of the which-way detector 
\begin{eqnarray}
|d_{\alpha}\rangle &=& {1\over\sqrt{3}}(|d_1\rangle + |d_2\rangle + |d_3\rangle)\nonumber\\
|d_{\beta}\rangle &=& {1\over\sqrt{3}}(e^{-i2\pi/3}|d_1\rangle + |d_2\rangle + e^{i2\pi/3}|d_3\rangle) \nonumber\\
|d_{\gamma}\rangle &=& {1\over\sqrt{3}}(e^{i2\pi/3}|d_1\rangle + |d_2\rangle + e^{-i2\pi/3}|d_3\rangle) .
\end{eqnarray}
In terms these states, (\ref{3ent1}) can be written as
\begin{eqnarray}
\Psi(x) &=& \tfrac{1}{3}\left[\psi_1(x)
+\psi_2(x)+ \psi_3(x)\right]|d_{\alpha}\rangle\nonumber\\
&&+ \tfrac{1}{3}\left[e^{-i2\pi/3}\psi_1(x) + \psi_2(x) + e^{i2\pi/3}\psi_3(x)\right]|d_{\beta}\rangle \nonumber\\
&& + \tfrac{1}{3}\left[e^{i2\pi/3}\psi_1(x) + \psi_2(x) + e^{-i2\pi/3}\psi_3(x)\right]|d_{\gamma}\rangle. \nonumber\\
\label{3ent2}
\end{eqnarray}
We assume that $\psi_1(x),\psi_2(x),\psi_3(x)$ are narrow states localized
at $x=-d,~x=0,~x=d$, respectively. We claim that (\ref{3ent2}) can also be
written in the following form
\begin{eqnarray}
\Psi(x) &=& \tfrac{1}{3}\left[\psi_1(x)
+\psi_2(x)+ \psi_3(x)\right]|d_{\alpha}\rangle\nonumber\\
&&+ \tfrac{1}{3}e^{ip_kx/\hbar}\left[\psi_1(x) + \psi_2(x) + \psi_3(x)\right]|d_{\beta}\rangle \nonumber\\
&&+ \tfrac{1}{3}e^{-ip_kx/\hbar}\left[\psi_1(x) + \psi_2(x) + \psi_3(x)\right]|d_{\gamma}\rangle , \nonumber\\
\label{3ent3}
\end{eqnarray}
where $p_k=2\pi\hbar/3d$ is a momentum kick. It is straightforward to see
that at $x=\pm d$, $e^{ip_kx/\hbar}= e^{\pm i2\pi/3}$, and (\ref{3ent3})
reduces to (\ref{3ent2}).

The state (\ref{3ent3}) implies that the particle passing through the triple
slit, passes undisturbed one third of the time (when which-way detector
state is $|d_{\alpha}\rangle$), experiences a momentum kick of magnitude
$p_k=h/3d$ one-third of the time (when the which-way detector
state is $|d_{\beta}\rangle$), and experiences a momentum kick of magnitude
$-h/3d$ one-third of the time (when the which-way detector
state is $|d_{\gamma}\rangle$). Thus the loss of interference in a 
three-slit interference experiment too, can be interpreted either as arising due to
entanglement of the three paths with three orthogonal which-way states,
or as arising due to the two kinds of momentum kicks the particle receives
at random.

\section{n-slit which-way experiment}

To complete the picture, we look at a n-slit interference
experiment with which-way detection, and enquire if the concept of
momentum kicks is general enough to be applicable to this case.
The state of the particle plus the which-way detector, as it emerges from
the n-slit, can be written as
\begin{equation}
	\Psi(x) = \tfrac{1}{\sqrt{n}} \sum_{k=1}^{n} \psi_k(x)|d_k\rangle ,
\label{nent1}
\end{equation}
where $\{|d_k\rangle\}$ are the n mutually orthogonal states of the
path-detector, and $|\psi_k\rangle$ is the state of the particle corresponding
to it passing through the k'th slit. The probability density of the particle,
at a position $x$ on the screen, is given by
\begin{eqnarray}
\Psi^*\Psi  = \tfrac{1}{n} \sum_{k=1}^{n} |\psi_k|^2
+ \tfrac{1}{n} \sum_{j,k} \psi_j^*\psi_k \langle d_j|d_k\rangle 
+ \psi_k^*\psi_j \langle d_k|d_j\rangle .
\label{nprob}
\end{eqnarray}
The interference, represented by the second summation, dies out because of
the orthogonality of $\{|d_k\rangle\}$.

Although the issue of wave-particle duality in n-slit interference has
recently been studied \cite{cd,nslit}, a quantum eraser has not been
theoretically formulated for n-slit interference. So, one has to
look for alternate basis states of the path-detector which are
unbiased with respect to $\{|d_j\rangle\}$. Such a 'Fourier basis' can
be constructed using the n'th roots of unity. Let the basis states be
given by $\{|\alpha_j\rangle\}$. They are related to $\{|d_j\rangle\}$
as follows:
\begin{eqnarray}
|d_1\rangle &=& \tfrac{1}{\sqrt{n}}\left(|\alpha_1\rangle 
	+ |\alpha_2\rangle + |\alpha_3\rangle
+ |\alpha_4\rangle + \dots + |\alpha_n\rangle\right)\nonumber\\
|d_2\rangle &=& \tfrac{1}{\sqrt{n}}\left(|\alpha_1\rangle 
+ e^{i2\pi\over n}|\alpha_2\rangle + e^{i4\pi\over n}|\alpha_3\rangle
	+ e^{i6\pi\over n}|\alpha_4\rangle + \right.\nonumber\\
&& \left. \dots + e^{i2(n-1)\pi\over n}|\alpha_n\rangle \right)\nonumber\\
|d_3\rangle &=& \tfrac{1}{\sqrt{n}}\left(|\alpha_1\rangle 
+ e^{i4\pi\over n}|\alpha_2\rangle + e^{i8\pi\over n}|\alpha_3\rangle
	+ e^{i12\pi\over n}|\alpha_4\rangle + \right.\nonumber\\
&& \left. \dots + e^{i4(n-1)\pi\over n}|\alpha_n\rangle \right) .
\end{eqnarray}
All the other states of $\{|d_j\rangle\}$ can be similary represented.
Elementary properties of n'th roots of unity ensure the orthonormality of
states. The entangled state (\ref{nent1}) can be represented in this new
basis as
\begin{eqnarray}
\Psi(x) &=& \tfrac{1}{n}\left(\psi_1 + \psi_2 + \psi_3 + \dots
+  \psi_n \right)|\alpha_1\rangle \nonumber\\
&&+ \tfrac{1}{n}\left(\psi_1
+ e^{i2\pi\over n}|\psi_2 + e^{i4\pi\over n}\psi_3
        + e^{i6\pi\over n}\psi_4 + \right.\nonumber\\
&& \left. \dots + e^{i2(n-1)\pi\over n}\psi_n \right)|\alpha_2\rangle\nonumber\\
&& + \tfrac{1}{n}\left(\psi_1
+ e^{i4\pi\over n}\psi_2 + e^{i8\pi\over n}\psi_3
        + e^{i12\pi\over n}\psi_4 + \right.\nonumber\\
&& \left. \dots + e^{i4(n-1)\pi\over n}\psi_n \right)|\alpha_3\rangle 
	+ \dots + \left(\dots\right) |\alpha_n\rangle . \nonumber\\
\end{eqnarray}
If the particle were to be detected in coincidence with the results of
measurement of an observable of which-way detector whose $n$ eigenstates
are $\{|\alpha_j\rangle\}$, each subset would yield a n-slit interference.
This would be a quantum eraser for n-slit interference. However, all 
the $n$ interference patterns, would be mutually shifted, and their
sum would wash out the interference.
Assuming that the slits are located at $x=0,d,2d,3d,\dots, (n-1)d$,
it can be shown that the above state can be written as
\begin{eqnarray}
\Psi(x) &=& \tfrac{1}{n}\left(\psi_1 + \psi_2 + \psi_3 + \dots
+  \psi_n \right)|\alpha_1\rangle \nonumber\\
&& + \tfrac{1}{n}e^{ip_1x\over\hbar}\left(\psi_1 + \psi_2 + \psi_3 + \dots
+  \psi_n \right)|\alpha_2\rangle \nonumber\\
&& + \tfrac{1}{n}e^{ip_2x\over\hbar}\left(\psi_1 + \psi_2 + \psi_3 + \dots
+  \psi_n \right)|\alpha_3\rangle \nonumber\\
	&& + \dots \nonumber\\
	&& + \tfrac{1}{n}e^{ip_{n-1}x\over\hbar}\left(\psi_1 + \psi_2 + \psi_3 + \dots
	+  \psi_n \right)|\alpha_{n}\rangle , \nonumber\\
\end{eqnarray}
where $p_j = jh/nd$. It should be asserted that in arriving at the above
result, one has used the fact that a state $\psi_j(x)$ is
a narrow state, virtually within the confines of a single slit, localized at
$x=(j-1)d$.
If the particle were to be detected in coincidence with
the states $\{|\alpha_j\rangle\}$, the particle state corresponding to
$|\alpha_1\rangle$ will be $\tfrac{1}{\sqrt{n}}\left(\psi_1 + \psi_2 + \psi_3
+ \dots +  \psi_n \right)$, which is the original state of the incoming 
particle, without any effect of the path-detector.
The particle state corresponding to another
path-detector state (say) $|\alpha_j\rangle$ will be the 
original state of the incoming particle, but with
an additional term $e^{ip_jx\over\hbar}$, which can be interpreted as
a momentum kick of magnitude $p_j = jh/nd$. 
So, the effect of introducing the which-way detection on the particle
can be interpreted as the particle either receiving no momentum kick, or
randomly receiving a kick of one of the n-1 magnitudes,
$p_j = jh/nd,~j=1, 2, \dots, n-1$. Largest kick is of magnitude $h/2d$
(for even $n$) and $\pm{n-1\over n}h/2d$ (for odd $n$), and the smallest momentum
kick (other than 0) is of magnitude $\pm h/nd$.

\section{Momentum space interference \& position kicks}

Recently Ivanov et al \cite{ivanov} proposed an interesting "double-slit"
experiment in momentum space. The basic idea is to consider particles in
superposition of two distinct momentum states. These distinct momentum
states play the role of distinct positions of the two slits in a conventional
two-slit experiment. They also proposed a which-way variant of their
proposed experiment. Let us see if the ideas in the preceding discussion
are applicable to this momentum space experiment. The state of a particle
in such a state, can be represented as
\begin{equation}
\Psi(p) = \tfrac{1}{\sqrt{2}} \left[ \psi_1(p) + \psi_2(p) \right],
\label{pent0}
\end{equation}
where $\psi_1(p)$ and $\psi_2(p)$ represent two states with distinct momenta
$p_1$ and $p_2$ respectively.
It was proposed that if the particle undergoes elastic scattering, it will
lead to interference \cite{ivanov}. However, we recognize that the source of
interference is the superposition state (\ref{pent0}). If now one introduces
a two-state which-way device, the combined state will look like the following
\begin{equation}
\Psi(p) = \tfrac{1}{\sqrt{2}} \left[ \psi_1(p)|d_1\rangle + \psi_2(p)|d_2\rangle \right],
\label{pent1}
\end{equation}
where $|d_1\rangle, |d_2\rangle$ are two orthogonal states of the which-way
detector. The entanglement with the which-way detector will lead to loss 
of interference.

Let us now consider this state in the basis described by $|d_+\rangle$ and
$|d_-\rangle$, introduced earlier. The state looks like the following
\begin{eqnarray}
\Psi(p) &=& \tfrac{1}{2} [ \psi_1(p)+ \psi_2(p)]|d_+\rangle 
 + \tfrac{1}{2} [ \psi_1(p) - \psi_2(p)]|d_-\rangle. \nonumber\\
\label{pent2}
\end{eqnarray}
The two states of the particle, corresponding to $|d_+\rangle$ and
$|d_-\rangle$, differ by a ``phase-flip."
Now an interesting observation is that this state can also be written as
\begin{eqnarray}
\Psi(p) &=& \tfrac{1}{2} [ \psi_1(p)+ \psi_2(p)]|d_+\rangle \nonumber\\
	&& + e^{-ip_1x_0/\hbar}e^{ipx_0/\hbar} \tfrac{1}{2} [ \psi_1(p) + \psi_2(p)]|d_-\rangle, \nonumber\\
\label{pent3}
\end{eqnarray}
where $x_0={h\over 2(p_2-p_1)}$. Notice that the first exponential factor
$e^{-ip_1x_0/\hbar}$ is just a constant phase factor. The second exponential
term $e^{ipx_0/\hbar}$ is such that it will just shift the position of 
a momentum eigenfunction by an amount $x_0$. Thus, this factor can treated
as a position kick the particle randomly receives, whenever the which-way
detector state is $|d_-\rangle$. Thus, the loss of interference in 
momentum space interference, due to the introduction of which-way detection,
can be attributed to random position kicks of magnitude ${h\over 2(p_2-p_1)}$.
It is interestingly analogous to  the momentum kicks of magnitude
$h/2d$ for position-space interference.

\section{Conclusion}

In conclusion, we have looked at the controvertial issue of momentum kick
that a particle might receive when passing through a double-slit in an
interference experiment involving which-way measurement. We have shown that there
are two completely equivalent ways of looking at the experiment. These two
ways correspond to two different basis sets of the which-way detector.
In one view,
the loss of interference is due to the entanglement of the two paths of the
particle to the two orthogonal states of the which-way detector. In another
view, the particle passing through the double-slit randomly receives a 
momentum kick of magnitude $h/2d$, $d$ being the slit separation. The
particles which receive a momentum kick, and those which do not receive a
kick, separately form two mutually shifted interference patterns, which
cancel each other when added. The same scenario holds for a three-slit 
interference pattern, except the particle either receives no kick, or
receives one of the two kinds of momentum kicks, of magnitude $\pm h/3d$.
The analysis has been generalized to the case of n-slit interference,
and it has been shown that the loss of interference can be interpreted
as the particle receiving either no kick, or one of the n-1 kinds of
kicks. It is shown that the analyses presented here applies equally well
to momentum space interference of Ivanov et.al. \cite{ivanov}, and the
loss of interference in a which-way measurement can be interpreted in terms
random {\em position kicks}.

In the analysis presented here, the momentum kick is experienced by the
particle passing through the slits, as it interacts with the which-way
detector, and it is not meaningful to ask if the momentum kick is received at
the location of the two slits or in between the two slits, a language
introduced by Wiseman \cite{wiseman2}.
As the two different views are based on exactly the same quantum state, the
two views are completely equivalent. However, it needs to be stressed that
what is common to both views is the entanglement between the particle paths
and the states of the which-way detector. This entanglement is at the root
of complementarity as, according to von Nuemann's first process of any quantum
measurement \cite{neumann}, any detector trying to determine which path the
particle followed, will necessarily get entangled with the particle
paths \cite{tqrv}. We believe this analysis should resolve any controversy 
regarding the momentum kicks in which-way experiments.

The author thanks Howard Wiseman for useful discussions.

\end{document}